\documentclass[aps,pre,twocolumn,groupedaddress,amsfonts,showpacs]{revtex4}
\usepackage{graphicx}

\usepackage[applemac]{inputenc}
\usepackage[T1]{fontenc}

\newcommand{\R}{{\bf R}}
\newcommand{\rr}{{\bf r}}
\newcommand{\G}{{\bf G}}
\newcommand{\thetaB}{\theta_{\rm B}}

\begin{document}
\title{Collective beating of artificial microcilia}
\author{Naïs Coq$^1$, Antoine Bricard$^1$, Francois-Damien Delapierre$^2$, Laurent Malaquin$^2$, Olivia du Roure$^1$, Marc Fermigier$^1$ and Denis Bartolo$^1$}
\affiliation{$^1$PMMH ESPCI-ParisTech-CNRS UMR 7636-Université Pierre et Marie Curie-Université Denis Diderot,10 rue Vauquelin 75231 Paris cedex 05 France.}
\affiliation{$^2$ Laboratoire Physico-Chimie Curie, UMR 168, Institut Curie, 11 rue Pierre et Marie Curie 75231 Paris cedex 05 France.}
\begin{abstract}
We combine technical, experimental and theoretical efforts to investigate the collective dynamics of artificial microcilia in a viscous fluid. We take advantage of soft-lithography and colloidal self-assembly to
devise microcapets made of hundreds of slender magnetic rods. This novel experimental setup  is used to investigate the dynamics of extended cilia
arrays driven by a precessing magnetic field.  Whereas the dynamics of an
isolated cilium is a rigid body rotation, collective beating results in a symmetry breaking of the
precession patterns. The trajectories of the cilia are anisotropic and experience a significant structural evolution as the actuation frequency increases. We present a minimal model to
account for our experimental findings and demonstrate how the global  geometry of the array imposes the shape of the trajectories via long range hydrodynamic interactions. 

\end{abstract}
\maketitle
Beating cilia are frequently encountered in nature to achieve propulsion, and to pump fluid at microscopic scales. 
Prominent examples include microorganisms, such as paramecia~\cite{Sleigh62}, algae, such as Volvox colonies~\cite{Kirk, drescher2010}, and ciliated epithelial tissues, which direct flow of mucosa and fluids over macroscopic scales, see e.g.~\cite{Sawamoto:2006p4096}. Recently, these propulsion mechanisms have sparked much interest in two distinct scientific communities. From a technological perspective, the past few years have witnessed the quick development of cilia-inspired microactuators, aimed at transporting and mixing fluids in microfluidic channels. These works focused mainly on the flow profile induced by the asymmetric actuation of the  artificial cilia~\cite{vilfan2010,superfine2010}. From a more fundamental perspective, the synchronization between beating cilia has motivated a surge in theoretical works~\cite{golestanianyeomans2011}, echoed by a few experimental studies on artificial~\cite{breuer2009} and biological setups~\cite{Polin:2009p4003}. So far, these experiments have been dedicated to two-body problems.  One of the main motivations for these works was to understand the role of the hydrodynamic coupling in the coordination of cilia beating in a viscous fluid. Most of the experiments and models have focused on the phase dynamics of these coupled active/actuated systems. Conversely, we report on the beating {\em amplitude} of actuated microcilia carpets. 
 
\begin{figure}[t]
  \center
\includegraphics[height=0.55\columnwidth]{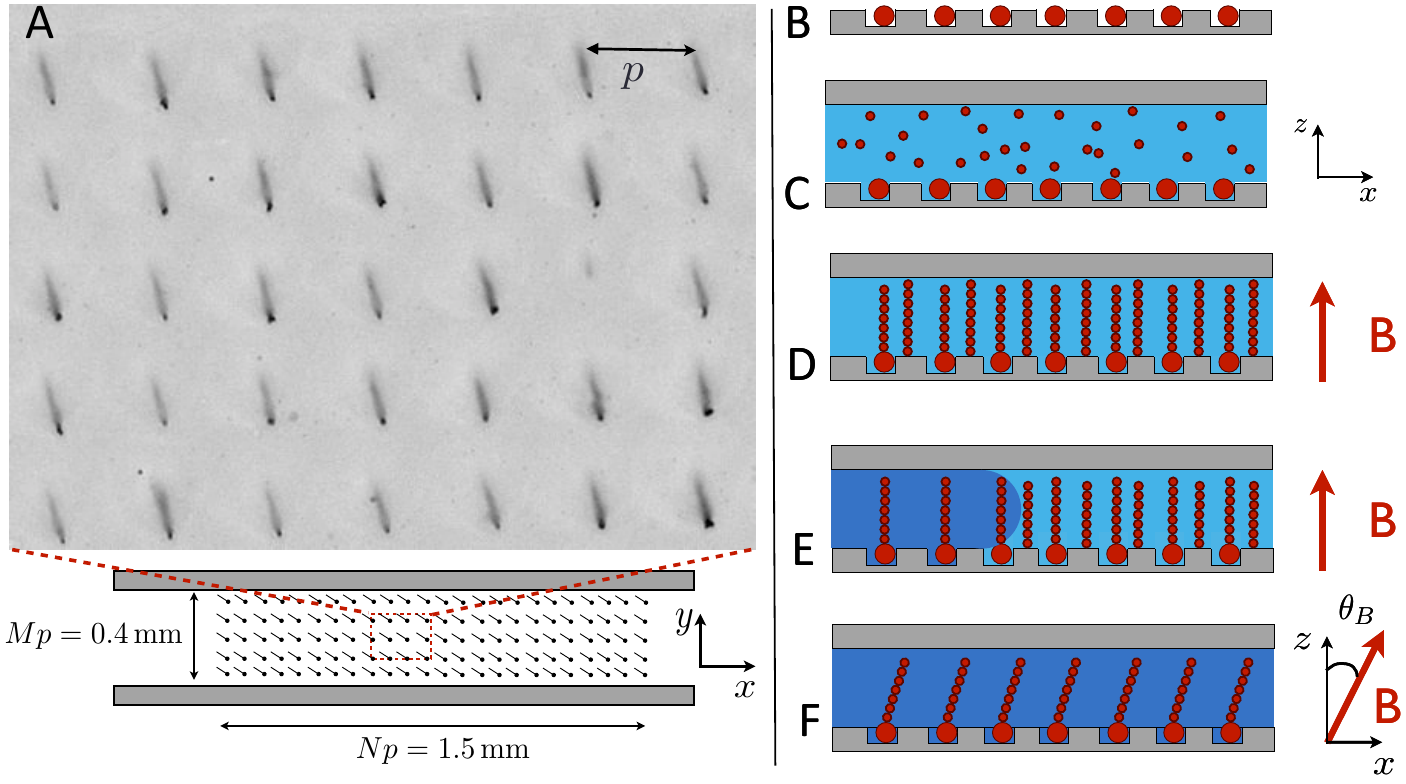}
 \caption{A: Snapshot of a square array of self-assembled colloidal filaments driven by a precessing magnetic field. The pitch of the array is $p=30\,\mu$m. The filaments are organized into a rectangular carpet inside a microchannel. B, C, D and E: 
 The filaments are formed by self-assembly of superparamagnetic colloids.
 A fraction of them aligns with magnetic anchoring sites at the bottom of the chamber. The excedentary filaments are rinsed away, leaving an array of filaments organized in the geometry imposed by the magnetic template.}
 \label{figure1}
 \end{figure}

\begin{figure}
  \center
    \includegraphics[width=1\columnwidth]{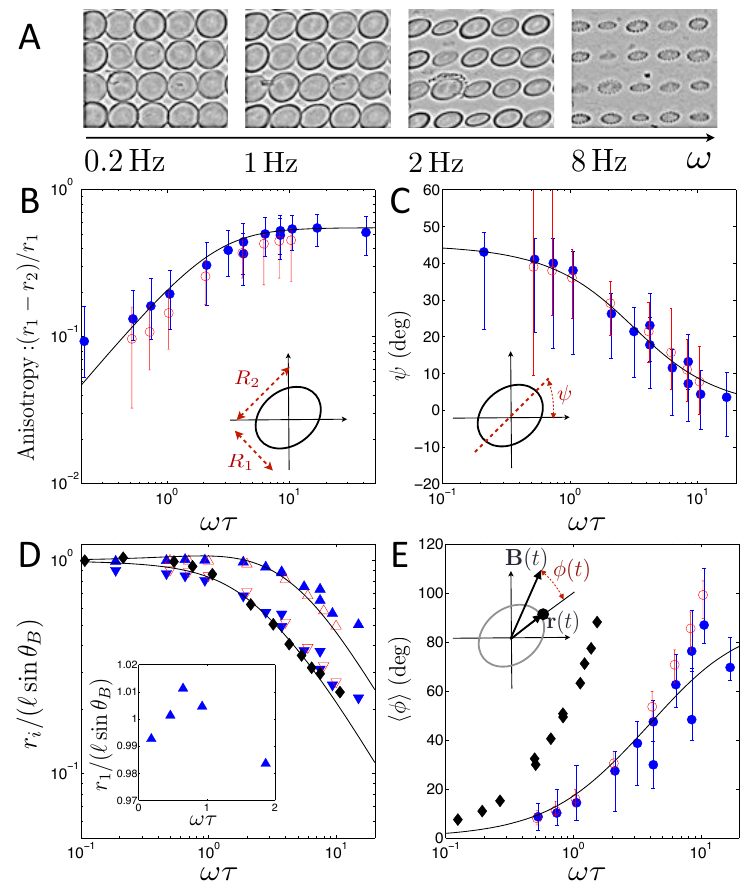} 
       \caption{A: Superimposed pictures revealing the trajectories followed by the  tip of the filaments. Lattice pitch: $30\,\rm\mu m$, field amplitude: $B=8\,$mT. B, C, D and E: symbols: experimental data. Full line: theoretical prediction form the mean field model. Blue filled symbols (resp. red open symbols) correspond to lattices with a  $p=40\,\mu \rm m$ pitch (resp. $p=30\,\mu \rm m$).  Field amplitudes: $B=4$ mt and $8$~mT.  B: Anisotropy as a function of the rescaled angular velocity. Straight line: Best fit from the mean-field model ($\tau_x=0.19\tau$, $\tau_y=0.43\tau$).  C: Orientation of the major axis as a function of the rescaled angular velocity.  D: Variations of the major axis and  minor axes with $\omega\tau$, and comparison with the variation of the trajectory radius for an isolated cilium (black diamonds). Inset: Close up on the low speed regime. The major axis displays non-monotonic variations. E: Variation of the time-averaged phase lag between the tip and the field orientation and  comparison with the same phase lag for an isolated magnetic cilium (black diamonds).}
    \label{figure2}
 \end{figure}

In this letter, we investigate  the dynamics of magnetic microrods driven  by an external precessing field in a viscous fluid, see Fig.~\ref{figure1}. We show that the long-range hydrodynamic interactions result in an unexpected symmetry breaking of the beating trajectories, Fig.~\ref{figure2}A. To address this many-body problem, we propose
 a novel design strategy for the fabrication of extended magnetic microcarpets, which we briefly describe below. In addition, we introduce a minimal but quantitative model to rationalize our experimental findings. It indicates that the shape and orientation of the trajectories reflect the large-scale geometry of the cilia array.   
 
{\it Magnetic microcilia carpets and experimental set-up.} Over the last four years, numerous fabrication strategies have been proposed to pattern microchannel surfaces with field-responsive magnetic microcilia arrays~\cite{dentoonder2008, superfine2010,vilfan2010}. However,  the accurate control of extended carpet geometries and the actuation via spatially homogeneous  fields have not been achieved simultaneously.  We combined soft lithography techniques and colloidal self-assembly to overcome these technical obstacles. By doing so, we managed to  make prototypal magnetic cilia arrays organized into tunable arrangements over large scales, Fig.~\ref{figure1}A.
The fabrication of the microfilaments themselves, via self-assembly of superparamagnetic colloids, has been previously  described~\cite{goubault2003}. However, the organization of these structures into controlled geometries remained virtually impossible. To achieve spatial ordering, we guide the self-assembly of colloidal chains, using a soft-lithographied magnetic template sketched in Fig.~\ref{figure1}B. In brief, using conventional replica molding techniques, we pattern a PDMS elastomer sheet with  square lattices of square holes (width: $6\,\mu$m, depth: $3\,\mu$m and pitch: $30\,\mu$m, or $40\,\mu$m). In each hole,  a single paramagnetic colloid (Dynabeads, diameter: $3.5\,\mu$m) is deposited via capillary assembly following~\cite{malaquin2007}. The resulting template, Fig.~\ref{figure1}B, is used to seal a PDMS microchannel (height: $150\,\mu$m, width: $400\,\mu$m, length 1 cm).  The channel is filled with colloidal particles of radius $a=375$~nm and magnetic susceptibility $\chi\sim 1$ (Ademtech). The colloids are diluted to a volume fraction $\Phi\sim 0,125\%$ in an aqueous solution containing $0.1$ wt$\%$ polyacrylic acid (Mw=250\,000, Sigma) and $0.1$ wt$\%$ nonyl phenol ethoxylate (surfactant NP10, $0.1$ wt$\%$, Sigma), Fig.~\ref{figure1}C. The flow is then stopped with pneumatic PDMS valves~\cite{galas2009}, and a 24 mT vertical magnetic field is applied immediately after the filling of the channel. The magnetic dipolar interactions between the paramagnetic particles cause them to organize into single stranded chains, aligned with the direction of the field and spread all over the channel, Fig.~\ref{figure1}D. The polyacrylic acid molecules adsorbed on their surface link the colloids irreversibly~\cite{goubault2003}.
 The excendentary filaments are subsequently rinsed away with a 50 wt$\%$ water-glycerol mixture (viscosity: $\eta=5\,$mPa.s), Fig.~\ref{figure1}E. The resulting microcilia carpet is actuated by  a set of three perpendicular Helmoltz coils, generating a 3D magnetic field, homogeneous  over the whole sample~\cite{avin10,coq2010}. The internal dipolar interactions along each filament tend to align its main direction with the magnetic field, see Fig.~\ref{figure1}F and supplementary movie. We filmed simultaneously $\sim 50$ cilia at 30 frames per second on an inverted Nikon microscope (TE2000) with a 20x objective,  Fig~\ref{figure1}A. Incidentally,note that  this prototyping technique allows in principle for a vast range of geometries, which can extend up to several millimeters.

{\it Collective beating of magnetic cilia.}
We consider the simplest istotropic 3D actuation cycle: the magnetic field precesses around the vertical axis at a constant angular velocity $\omega$, keeping a constant angle $\theta_{\rm B}=15^\circ$ with the $z$-axis, Fig.~\ref{figure1}E. We first recall that in this small inclination limit, an isolated cilia responds linearly to the driving field, and the position of its tip 
follows a circular trajectory at a constant angular speed $\omega$~\cite{coq2010}. This is not what is observed with the cilia carpets. All the cilia do move synchronously at a constant angular speed $\omega$, but the tip trajectories break the rotational symmetry of the driving, as shown on the pictures in Fig.~\ref{figure2}A. The increase in the driving frequency results in the stretching of the beating patterns. In addition, the angle, $\psi$, between the main axis of these anisotropic trajectories and the $x$-axis decreases with $\omega$ (the $x$-axis is here defined by the channel orientation). Changing the sign of the actuation results in  mirrored trajectories.  In order to quantify these structural changes, we fitted the trajectories by ellipses, and plotted the variation of the anisotropy, and of the orientation with $\omega$ in Figs.~\ref{figure2}B and C. We defined the anisotropy as  $(r_1-r_2)/r_1$, where $r_1$ and $r_2$ are respectively  the major and the minor axis of the ellipses. The dispersion of the data mainly originates from the  polydispersity of the filament lengths: $\ell=135\,\mu{\rm m}\pm10\,\mu{\rm m}$. The error bars correspond to the maximal deviation from the mean values. Before going further, let us introduce the orientational relaxation time, $\tau$,  of an isolated rod. $\tau$ is defined  by the force balance between the magnetic and viscous forces acting on the cilia upon a sudden change in the field orientation. These forces scale as: $f_{\rm mag}\sim(a\chi B)^2/(\mu_0\ell)$ and $f_{\rm v}\sim\zeta_\perp (\ell/\tau)$ respectively, where  $\zeta_\perp\equiv4\pi\eta/[\ln(\ell/a)+1/2]$ is the rod friction coefficient in a slender body approximation\cite{Brennen:1977p4607}. We measured $\tau$ according to the procedure introduced in~\cite{coq2010}, which yields $\tau$=$(4.37\, 10^{-3})/B^2$ for our experiments, where the length of the filaments has been kept constant. $\tau$ is typically of the order of a few seconds. 
ind

First, we note that  all the measurements obtained for different field amplitudes, $B$, collapse on the same master-curves when we rescale the angular speed $\omega$ by $\tau$, Figs.~\ref{figure2}B,C, D and E. This implies that $\omega\tau$, the so-called Mason number, is the only dimensionless parameter involving the driving speed in this problem.  At low speed, $\omega\tau\ll1$, the anisotropy increases linearly with $\omega$, and $\psi$ slowly decays from a finite value close to $45^\circ$. At high speed, $\omega\tau\gg1$, the anisotropy plateaus to a value close to $0.5$ and the trajectories align with the $x-$axis. In addition, we compare  the major and minor axes individually to the radius, $r_0$, of the circular trajectory followed by an isolated cilium, see Fig.~\ref{figure2}D. In fact, the variations of $r_0$ and $r_2$ cannot be distinguished: they both start from $\ell\sin\theta_B$ and decay to 0 above a well-defined crossover at $\omega\tau=1$. Conversely, $r_1$, remains close to the static value $\ell\sin\theta_B$ up to $\omega\tau\sim5$, and then decays to 0 with the same slope as $r_2$. Looking more closely at the low speed regime, we notice that $r_1$ undergoes non-monotonic variations and reaches its maximal amplitude at $\omega\tau\sim 1$, see the inset in Fig.~\ref{figure2}D. These observations reveal that the cilia have more than  one relaxation time when beating within a carpet. At this point, we can anticipate on our theoretical model and infer that those two relaxation times arise from geometrical corrections to $\tau$, since 
 all the data collapse with $\omega\tau$. 
 To close this overview, we point that the time-averaged phase lag, $\langle\phi\rangle$, between the rod and the field orientation is significantly reduced compared to the case of a single cilium. Nonetheless, the shape of the curve $\langle \phi \rangle(\omega\tau)$ is conserved, see Fig.~\ref{figure2}E.

\begin{figure}
  \center
    \includegraphics[width=\columnwidth]{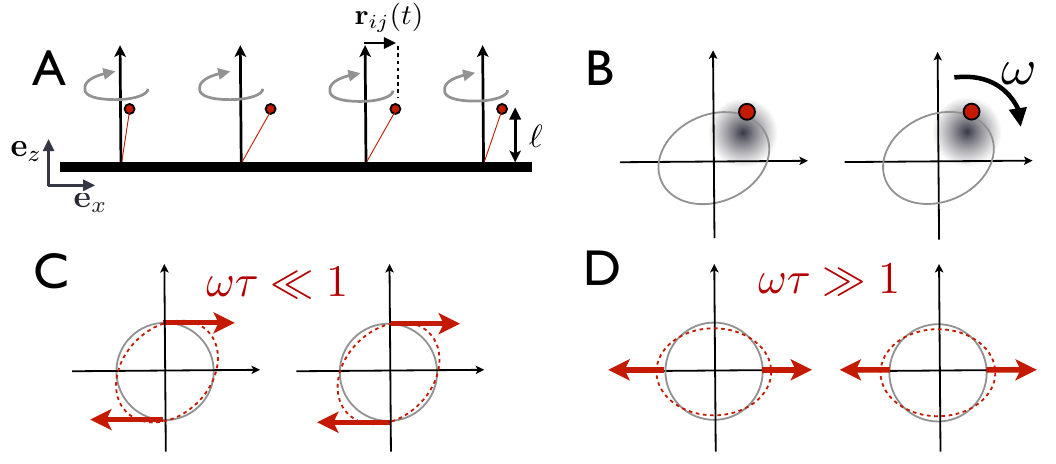}
         \caption{A: Sketch of geometry used in the theoretical model. B: Two synchronous rotators driven by a harmonic trap. C: Direction of the induced forces in the low-speed limit. The trajectories are effectively sheared by the hydrodynamic coupling. D: Same picture in the high-speed regime}
    \label{figure3}
 \end{figure}

{\it Theoretical model and physical interpretations.} 
To explain quantitatively the symmetry breaking of the trajectories, we introduce a minimal far-field model. The rectangular cilia carpet is modeled by a $2N\times 2M$  lattice of pointwise particles located at $\R_{ij}=p(i\,{\bf e}_x+j\,{\bf e}_y)+{\bf r}_{ij}(t)$, at a distance $\ell$ from a solid wall, see Fig.~\ref{figure3}. Each particle is driven by an external time-dependent  force ${\bf f}(\R_{ij},t)$. 
Note that we disregard the effect of the magnetic coupling between the cilia. 
The velocity of the particle $(i,j)$ is related to the force acting on all the other particles via: 
\begin{equation}
\partial_t \R_{ij}=\sum_{n,m}\G(\R_{ij}-\R_{n,m})\cdot{\bf f}(\R_{n,m},t),
\label{Eq1}
\end{equation}
where, for sake of simplicity, the hydrodynamic coupling between the particles is described by the Blake-Oseen tensor $\G$. $\G$ is the Green function of the Stokes equation associated to a force monopole oriented parallel to a solid wall in a viscous fluid ~\cite{blake71}. By definition, $\G(0)\equiv\zeta^{-1}{\mathbb I}$, is the isotropic mobility tensor for an isolated particle. 
We now perform a mean field approximation and assume that all the cilia follow identical trajectories in a synchronous manner: $\rr_{ij}(t)=\rr(t)$.  This approximation is justified by our experiments: we did not observe any spatial heterogeneities in the phase of the cilia tips.
Assuming that $\ell>p$,  at leading order in $p/\ell$, we have $\G(\R)=\frac{3\ell^2\cos^2\thetaB}{2\pi\eta R^5}{\R\R}$, for $\R\neq0$. After some algebra, Eq.\ref{Eq1} then reduces  to the equation of motion for a single anisotropic particle driven by a time dependent force:
\begin{eqnarray}
\partial_t\rr&=&\left(\alpha  \mathbb I+\beta {\bf e}_x{\bf e}_x\right)\cdot {\bf f}(\rr,t),\label{Eq2}
\end{eqnarray}
where  the two effective mobility coefficients are given by:
\begin{eqnarray}
\alpha&=&\zeta^{-1}+\frac{3\ell^2}{2\pi\eta p^3}\sum_{n=-N}^N\sum_{m=-M}^M\frac{m^2}{(n^2+m^2)^{5/2}}\label{alpha},\\
\beta&=&\frac{3\ell^2}{2\pi\eta p^3}\sum_{|n|>M}\sum_{m=-M}^M\frac{n^2-m^2}{(n^2+m^2)^{5/2}}\label{beta}.
\end{eqnarray}
To compute the trajectories, we now need to specify ${\bf f}(t)$.
As we consider only  small field inclinations, the magnetic actuation is well approximated by a rotating harmonic trap: ${\bf f}(t)=\nabla k\left[\rr-{\bf r}_0(t)\right]^2$, with ${\bf r}_0(t)=\ell\sin\theta_{\rm B}\left(\cos \omega t \,{\bf e}_x+\sin \omega t \,{\bf e}_y\right)$~\cite{coq2010}. Note that relaxing this harmonic hypothesis for the force and the geometric condition $\ell>p $ does not qualitatively change our predictions, as Eq.~\ref{Eq2} retains the same form. In the present case Eq.~\ref{Eq2} corresponds to the equation of motion for two uncoupled 1D overdamped harmonic oscillators driven by sinusoidal forces. They are readily solved, and the beating trajectories $\rr(t)\equiv[x(t),y(t)]$ are:
\begin{eqnarray}
x(t)&=&\frac{\ell\sin\theta_B}{\sqrt{1+(\omega\tau_x)^2}}\cos\left[\omega t+\arctan(\omega\tau_x)\right]\label{x}\\
y(t)&=&\frac{\ell\sin\theta_B}{\sqrt{1+(\omega\tau_y)^2}}\sin\left[\omega t+\arctan(\omega\tau_y)\right]\label{y}
\end{eqnarray} 
where $\tau_x\equiv 1/[k(\alpha+\beta)]$ and $\tau_y\equiv 1/ (k\alpha)$ are the two relaxation times of the particles, along the $x$ and $y$ directions respectively. Defining the major and the minor axes  as the extrema of $r(t)$, we can compute numerically $r_i(\omega)$ for $i=1,2$, $\psi(\omega)$ and $\Delta\phi(\omega)$. To 
 test our theoretical predictions, we first determined the two free parameters $\tau_x$ and $\tau_y$ by fitting the anisotropy data, Fig.~\ref{figure2}B, and then used the same parameters to calculate the major/minor axes, the inclination and the phase lag. As shown in Fig.~\ref{figure2}, our model yields excellent agreement with the experiments. This unambiguously proves that the anisotropic trajectories originate only from the hydrodynamic coupling between the cilia, and not from their magnetic interactions. 
 
 Moreover, Eqs.~\ref{x} and~\ref{y} imply that the symmetry breaking of the trajectories reflects  the large scale anisotropy of the cilia carpet. The trajectories are anisotropic if $\tau_x\neq\tau_y$, or equivalently if $\beta\neq0$. This mobility coefficient vanishes for symmetric (square) carpets only: indeed it arises from the interactions between particles located at a distance larger than the width of the cilia carpet, see Eq.\ref{beta}. This coupling enhances the particle mobility along the $x$-direction, thereby effectively stretching the beating trajectory. 
 
More quantitatively, the minor and the major axes are reached at $t^*$, defined as $\partial_t r^2(t)|_{t^*}=0$. At low speed, expanding Eqs.~\ref{x} and \ref{y} at first order in $\omega\tau$, we obtain $\partial_tr^2(t)=-2\omega(\tau_x/\tau_y)\omega\tau\cos(2\omega t)$, which implies that the trajectories are oriented at $\omega t^*=45^\circ$. Interestingly,  this orientation is independent of the two relaxation time values, provided that $\tau_x\neq\tau_y$. We can provide more physical  insight into this purely geometric result by looking at the structure of Eq.~\ref{Eq2}. The equation has the same structure for all carpet sizes and aspect ratio: understanding the two-body problem is sufficient to account for the structural change of the beating patterns. Eq.~\ref{Eq2} can be thought of as the equation of motion for two hydrodynamically coupled synchronous particles, see Fig.~\ref{figure3}B. When $\omega\tau\ll1$, the particle closely follows the trap motion, and the force acting on the cilia is tangent to the trajectory. Therefore, the force induced by the rotation of the second cilia on the first one, $\beta({\bf f}\cdot{\bf e}_x){\bf e}_x$, is extremal for $\omega t=\frac{\pi}{2}\,{\rm mod}(\pi)$, and vanishes for $\omega t=0\,{\rm mod}(\pi)$. This results in a net shear of the initially circular trajectory, as depicted in Fig.~\ref{figure3}C. Consequently, the trajectories are expected to be stretched at a $45^\circ$ angle from the $x$-axis, which is precisely what we observe in our experiments, Fig.~\ref{figure2}C. We emphasize that this last picture is generic, and does not depend on the specifics of the driving force. Since the amplitude  of the induced force scales as $r\omega$, the stretching of the trajectory is expected to grow linearly with the driving speed. Again this qualitative prediction is in agreement with our measurements and our asymptotic analysis. Indeed, estimating $x$ and $y$ at $t^*=\pi/(4\omega)$, we obtain $r_i=1-(-1)^{i}(\tau_x/\tau_y)(\omega\tau)$, with $i=1,2$. We thus correctly predict that the major axis increases  with the driving frequency, while the minor axis decreases, Fig.\ref{figure2} D and inset. As a result  the corresponding anisotropy $(r_1-r_2)/r_1=(\tau_x/\tau_y)(\omega\tau)$ increases linearly with  $\omega$, in agreement with our experimental data shown in Fig.~\ref{figure2}B.
 
In the opposite high speed limit, $\omega \tau\gg1$, the phase lag of the driven overdamped particles reaches $\pi/2$: the positions at which the induced forces are extremal are rotated by a $\pi/2$ angle with respect to the low-speed value. Therefore, as sketched in Fig.~\ref{figure3}D, the hydrodynamic coupling results in a net stretching of the trajectories along the $x$-axis. This simple picture is confirmed by Eqs.\ref{x} and \ref{y}. For $\omega\tau\gg1$, they reduce to the parametrization of an elliptic trajectory aligned with the $x$-axis. Both $r_1$ and $r_2$ decrease as $1/\omega$, since the particle is driven faster than it can respond to the field: $r_1\sim\ell\sin\theta_B/(\omega\tau_x)$ and  $r_2\sim\ell\sin\theta_B/(\omega\tau_x)$.  For $\omega\tau\gg1$, $\psi$ decays to 0, and the anisotropy plateaus to a constant value given by the ratio between the two relaxation times, $\tau_x/\tau_y$. This again confirms our experimental findings, Fig.\ref{figure2}B and C. In addition, $r_1$ increases for  $\omega\tau\ll1$, whereas it decreases for  $\omega\tau\gg 1$, reaching a maximal value close to $\omega\tau=1$ as shown in the inset in Fig.\ref{figure2}.

In conclusion, we combined technical, experimental and theoretical efforts to investigate the collective dynamics of actuated microcilia in a viscous fluid. By doing so, we uncovered unexpected morphologies in the beating trajectories. We showed that  this counter-intuitive dynamic response is a consequence of the long range hydrodynamic interactions. Importantly, we found  the shape of the precession patterns to be chiefly selected by the large scale geometry of the carpets.

We thank Sandrine Ngo for fruitful interactions, and Eric Lauga for a critical reading of the manuscript.
We acknowledge support by C'Nano IdF, Sesame Ile de France  and Paris \'emergence. 

\end{document}